\documentclass[aps,pra,amsmath,amssymb,showpacs, twocolumn]{revtex4-1}

\usepackage[T1]{fontenc}
\usepackage{amssymb,amsmath}
\usepackage{amsfonts}

\usepackage{mathbbol}
\usepackage{bbm}
\usepackage{dsfont}

\usepackage{graphicx,epsfig,epsf}

\def\ket#1{|#1\rangle}
\def\bra#1{\langle#1|}

\begin{document}

\title{Artificial abelian gauge potentials induced by dipole-dipole interactions between Rydberg atoms}
\author{A.~Cesa}
\author{J.~Martin}
\email{jmartin@ulg.ac.be}
\affiliation{Institut de Physique Nucl\'eaire, Atomique et de Spectroscopie, Universit\'e de Li\`ege, B\^at.\ B15, B - 4000 Li\`ege, Belgium}
\date{\today}
\begin{abstract}
We analyze the influence of dipole-dipole interactions between Rydberg atoms on the generation of Abelian artificial gauge potentials and fields. When two Rydberg atoms are driven by a uniform laser field, we show that the combined atom-atom and atom-field interactions give rise to new, nonuniform, artificial gauge potentials. We identify the mechanism responsible for the emergence of these gauge potentials. Analytical expressions for the latter indicate that the strongest artificial magnetic fields are reached in the regime intermediate between the dipole blockade regime and the regime in which the atoms are sufficiently far apart such that atom-light interaction dominates over atom-atom interactions. We discuss the differences and similarities of artificial gauge fields originating from resonant dipole-dipole and van der Waals interactions. We also give an estimation of experimentally attainable artificial magnetic fields resulting from this mechanism and we discuss their detection through the deflection of the atomic motion.
\end{abstract}


\maketitle

\section{Introduction}

In 1996, Lloyd~\cite{Llo96} showed that the dynamics of complex many-body quantum systems can be efficiently simulated by quantum computers, an idea first put forward by Manin~\cite{Man80} and further developed by Feynman~\cite{Fey82}. Although the first quantum computers of a few qubits have been realized experimentally~\cite{DiC09,Xu12}, the advent of scalable quantum computers might take another few decades. An alternative tool in the context of simulation is a highly controllable quantum system able to mimic the dynamics of other complex quantum systems, known as an analog quantum simulator. Cold neutral atoms and trapped ions have been shown to be versatile quantum simulators~\cite{Bul09,Blo12} thanks to their high flexibility, controllability, and scalability. They permit one to study a wide range of problems arising from atomic physics, relativistic quantum physics, or cosmology~\cite{Bla12}. Since neutral atoms do not carry any net charge, the simulation of electric and magnetic condensed matter phenomena, such as the spin Hall effect, seems out of reach. To overcome this apparent difficulty, the idea has been proposed to create artificial electromagnetic potentials for neutral atoms based on atom-light interaction~\cite{Juz04,Juz05,Juz06,Dal11}. These artificial  potentials act on neutral atoms as real electromagnetic potentials act on charged particles. Soon after, proposals for the generation of non-Abelian gauge potentials came out~\cite{Una99,Ost05,Rus05}, inspired by the work of Wilczek and Zee~\cite{Wil84} on adiabatic evolution in the presence of degenerate eigenstates. 

Many works on artificial gauge potentials induced by atom-light interaction adopt a single-particle approach. The predicted potentials are then supposed to be valid for a system of many weakly interacting atoms, like, e.g., in a Bose-Einstein condensate~\cite{Lin09a,Lin09b,Lin11,Bee13}. However, new physics arise in ultracold dipolar gases where long-range interactions between atoms are known to play an important role~\cite{Stu05,Dau12,Gru13}. So far, the consequences of atom-atom interactions on the generation of artificial gauge fields has little been studied. In particular, dipole-dipole interactions could significantly modify artificial gauge fields, or even give rise to new artificial gauge fields otherwise not present. The aim of this paper is to address this problem analytically by studying a system of two long-lived interacting Rydberg atoms driven by a common laser field. In a recent work~\cite{Kif13}, another configuration was studied in which two interacting Rydberg atoms are subjected to a static electric field. Abelian and non-Abelian artificial gauge fields were computed numerically for different amounts of Stark shifts asymmetry of the Zeeman sublevels, and were shown to realize a velocity-dependent beam splitter~\cite{Kif13b}.

The paper is organized as follows. In Sec.~II, we present some established results on artificial gauge fields for a single two-level atom evolving adiabatically~\cite{Dal11} as these will be used throughout the paper. We give at the same time a different formulation of the single-atom artificial gauge potentials. We then generalize the analysis to two noninteracting two-level atoms. In Sec.~III, we study a system of two interacting Rydberg atoms driven by a uniform laser field and calculate the associated artificial gauge potentials when the system evolves adiabatically. We discuss the general expressions and explain the mechanism responsible for these potentials and fields. We estimate the attainable field strengths and consider their detection in view of recent experiments. A brief conclusion is drawn in Sec. IV. Sections V and VI are appendixes dedicated to the calculation of the artificial gauge potentials in the center-of-mass coordinate system and to the derivation of an effective Hamiltonian in the dipole blockade regime.

\section{Artificial gauge potentials without atom-atom interactions}
\subsection{A single two-level atom}
We consider a single two-level atom interacting with a classical electric field $\mathbf{E}(\mathbf{r},t)=\mathcal{E}_0 \,\boldsymbol{\epsilon} \cos(\mathbf{k}_L\boldsymbol{\cdot}\mathbf{r}-\omega_L t)$ of amplitude $\mathcal{E}_0$, polarization $\boldsymbol{\epsilon}$, and wavevector $\mathbf{k}_L$. In the electric dipole and rotating wave approximations, the Hamiltonian accounting for the internal dynamics of the atom is given, in the interaction picture, by
\begin{equation}\label{H2l}
\hat{H}_{2l}= \frac{\hbar}{2} \left(\Omega\,e^{i\mathbf{k}_L\boldsymbol{\cdot}\mathbf{r}} |e\rangle\langle g|+\mathrm{H.c.}\right)-\frac{\hbar\delta}{2}\sigma_z
\end{equation}
where $\ket{e}$, $\ket{g}$ are the atomic excited and ground states respectively, separated in energy by $\hbar\omega$, $\Omega =-\mathbf{d}^*\boldsymbol{\cdot}\boldsymbol{\epsilon}\,\mathcal{E}_0/\hbar$ is the Rabi frequency with $\mathbf{d}=\langle g|\hat{\mathbf{d}}|e\rangle$ the dipole matrix element for the transition $\ket{e}\leftrightarrow\ket{g}$, $\delta=\omega_L-\omega$ is the detuning between the laser and atomic transition frequencies, $\mathrm{H.c.}$ stands for Hermitian conjugate and $\sigma_z=|e\rangle\langle e|-|g\rangle\langle g|$. Hamiltonian (\ref{H2l}) admits two nondegenerate eigenvectors:
\begin{equation}\label{chipm}
\ket{\chi_\pm(\mathbf{r})}=\frac{\left(-\delta \pm\Lambda\right)e^{i\mathbf{k}_L\boldsymbol{\cdot}\mathbf{r}}}{\sqrt{2\left(\Lambda^2\mp\delta\Lambda\right)}}\,\ket{e}+\frac{\Omega^*}{\sqrt{2\left(\Lambda^2\mp\delta\Lambda\right)}}\,\ket{g}
\end{equation}
of energy $E_\pm=\pm\hbar\Lambda/2$ where 
\begin{equation}\label{Lambda}
\Lambda\equiv\Lambda(\mathbf{r})=\sqrt{|\Omega(\mathbf{r})|^2+\delta(\mathbf{r})^2}
\end{equation} 
is the generalized Rabi frequency. The latter can vary with the atomic position both through the Rabi frequency $\Omega$ via the electric field amplitude and through the detuning which can be made position dependent by means of the Zeeman effect~\cite{Lin09b,Lin11}.

When the atomic motion is treated quantum-mechanically, the Hamiltonian governing the system is given in the position representation by
\begin{equation}\label{Hamfull}
\hat{H}=\left(\frac{\hat{\mathbf{p}}^2}{2m}+U\right)\otimes \hat{\mathbb{1}}^{\mathrm{int}}+\hat{\mathbb{1}}^{\mathrm{ext}}\otimes \hat{H}_{2l}
\end{equation}
where $\hat{\mathbf{p}}=\hbar\boldsymbol{\nabla}_{\mathbf{r}}/i$ is the atomic momentum operator, $m$ is the atomic mass, $U$ is a potential energy insensitive to the atomic internal state, $\hat{H}_{2l}$, given by Eq.~(\ref{H2l}), is the atomic internal Hamiltonian, and $\hat{\mathbb{1}}^{\mathrm{int}}$ ($\hat{\mathbb{1}}^{\mathrm{ext}}$)  denotes the identity operator in the atomic internal (external) Hilbert space. 

The internal state of the atom can always be decomposed onto the basis states (\ref{chipm}) which depend parametrically on the atomic position $\mathbf{r}$. With such a decomposition, the global wave function of the atom in the position representation reads~\cite{Dal11}
\begin{equation}
\langle\mathbf{r}|\psi(t)\rangle= \sum_{j=\pm}{\psi_j(\mathbf{r},t) |\chi_j(\mathbf{r})\rangle}.
\end{equation}

When the atom is initially in the internal state $|\chi_+(\mathbf{r})\rangle$ and moves sufficiently slowly to ensure adiabatic evolution, it stays over time in the internal state $|\chi_+(\mathbf{r})\rangle$. In this case, the population of the state $|\chi_-(\mathbf{r})\rangle$ remains negligible such that $\langle\mathbf{r}|\psi(t)\rangle\approx \psi_+(\mathbf{r},t) |\chi_+(\mathbf{r})\rangle$ at any time $t$. Plugging this expression of the wave function into the time-dependent Schr\"{o}dinger equation for Hamiltonian (\ref{Hamfull}) and projecting onto $|\chi_+(\mathbf{r})\rangle$, we obtain
a Schr\"{o}dinger-like equation for $\psi_+(\mathbf{r},t)$,
\begin{equation}\label{multi2levHeff}
\begin{aligned}
i\hbar\frac{\partial }{\partial t}\psi_+(\mathbf{r},t)=\bigg[& \frac{\big(\hat{\mathbf{p}}-q\mathbf{A}^+\big)^2}{2m}+q\phi^++U+\frac{\hbar \Lambda}{2}\bigg]\psi_+(\mathbf{r},t),
\end{aligned}
\end{equation}
with
\begin{equation}
q\mathbf{A}^+(\mathbf{r}) =i\hbar\langle \chi_+|\boldsymbol{\nabla}_{\mathbf{r}}\chi_+\rangle,
\label{2nivA}
\end{equation}
and
\begin{equation}\label{phieff}
q\phi^+(\mathbf{r})=\frac{\hbar^2}{2m}|\langle\chi_-|\boldsymbol{\nabla}_{\mathbf{r}}\chi_+\rangle|^2.
\end{equation}

Equation (\ref{multi2levHeff}) is formally equivalent to Schr\"{o}dinger's equation for a particle of electric charge $q$ immersed in an electromagnetic field described by the potentials $\mathbf{A}^+(\mathbf{r})$ and $\phi^+(\mathbf{r})$ and experiencing an additionnal potential $U+\hbar \Lambda/2$. The potentials $\mathbf{A}^+(\mathbf{r})$ and $\phi^+(\mathbf{r})$ are therefore referred to as Abelian artificial gauge potentials. The completeness relation $\ket{\chi_+}\bra{\chi_+}+\ket{\chi_-}\bra{\chi_-}=\hat{\mathbb{1}}^{\mathrm{int}}$ allows us to rewrite Eqs.~(\ref{2nivA}) and (\ref{phieff}) in terms of the expectation value $\langle\hat{\mathbf{p}}\rangle_{\chi_+}$ and variance $(\Delta \hat{\mathbf{p}}^2)_{\chi_+}$ of the momentum operator $\hat{\mathbf{p}}=\hbar\boldsymbol{\nabla}_{\mathbf{r}}/i$ in the position-dependent internal state $\ket{\chi_+(\mathbf{r})}$, 
\begin{equation}
q\mathbf{A}^+(\mathbf{r}) =-\langle\hat{\mathbf{p}}\rangle_{\chi_+},\;\;\;\;\;q\phi^+(\mathbf{r})=\frac{(\Delta \hat{\mathbf{p}}^2)_{\chi_+}}{2m}.
\label{2nivAb}
\end{equation}
This formulation of the artificial gauge potentials makes explicit the interpretation of the term $\big(\hat{\mathbf{p}}+\langle\hat{\mathbf{p}}\rangle_{\chi_+}\big)^2/2m$ appearing in (\ref{multi2levHeff}) as the kinetic energy associated with the slow center-of-mass motion of the atom during its adiabatic evolution. It also makes clear that the origin of the scalar potential $\phi^+$ lies in the quantum fluctuations of momentum as measured by the variance, in full agreement with the interpretation of this term as additional kinetic energy associated with the micro-motion of the atom resulting from its interaction with the laser field~\cite{Che08}.

In the remainder of this paper, we set the artificial charge $q$ equal to one unless otherwise stated. When the Rabi frequency has a constant phase $\varphi$, $\Omega(\mathbf{r})=|\Omega(\mathbf{r})|e^{i\varphi}$, we directly obtain from Eqs.~(\ref{chipm}), (\ref{2nivA}), and (\ref{phieff}),
\begin{equation}
\begin{aligned}
\mathbf{A}^+(\mathbf{r}) ={}&\left(-1+\frac{\delta}{\Lambda}\right)\frac{\hbar \mathbf{k}_L}{2},\\
\phi^+(\mathbf{r})={}&\bigg[\frac{(\delta\, |\boldsymbol{\nabla}_{\mathbf{r}}\Omega|+|\Omega|\, |\boldsymbol{\nabla}_{\mathbf{r}}\delta|)^2}{k_L^2\Lambda^4}+\frac{|\Omega|^2 }{\Lambda^2}\bigg]\frac{\hbar^2k_L^2}{8m}.
\end{aligned}
\label{2leva-phi-}
\end{equation}
If the system had adiabatically followed the state $\ket{\chi_-}$ (instead of $\ket{\chi_+}$), the resulting scalar potential would be identical to Eq.~(\ref{2leva-phi-}) whereas the vector potential would appear with the opposite sign in front of $\delta/\Lambda$.

The artificial magnetic and electric fields associated with these potentials are $\mathbf{B}^\pm=\mathbf{\nabla}_{\mathbf{r}}\times\mathbf{A}^\pm$ and $\mathbf{E}^\pm=-\mathbf{\nabla}_{\mathbf{r}}\phi^\pm$. For instance, the artificial magnetic field is given by
\begin{equation}
\mathbf{B}^\pm(\mathbf{r})=\pm\frac{\hbar |\Omega|}{2\Lambda^3} \big(|\Omega|\,\boldsymbol{\nabla}_{\mathbf{r}}\delta-\delta\,\boldsymbol{\nabla}_{\mathbf{r}}|\Omega|\big)\times \mathbf{k}_L.
\label{2levBart}
\end{equation}
Whenever $\delta$ and $\Omega$ are uniform, i.e., do not vary in space, the artificial vector potential is uniform and the magnetic field vanishes everywhere.

\subsection{Two noninteracting two-level atoms}

For comparison with latter results and to introduce notations, we briefly consider the case of two noninteracting two-level atoms driven by a common laser field. Let us denote by $\mathbf{r}_\alpha$ the position of atom $\alpha=a,b$. The Hamiltonian describing the internal dynamics of the atoms interacting with the laser field is given by
\begin{equation}\label{Hind}
\hat{H}_{\mathrm{non-int}} = \hat{H}_{2l,a}\otimes \hat{\mathbb{1}}_b^{\mathrm{int}} +\hat{\mathbb{1}}_a^{\mathrm{int}}\otimes \hat{H}_{2l,b},
\end{equation}
where $\hat{\mathbb{1}}_\alpha^{\mathrm{int}}$ denotes the identity operator in the internal Hilbert space of atom $\alpha$ and $ \hat{H}_{2l,\alpha}$ is the single-atom interaction Hamiltonian (\ref{H2l}) for atom $\alpha$ ($\alpha=a,b$). The eigenvectors of Hamiltonian (\ref{Hind}) follow directly from the eigenvectors $\ket{\chi_\pm(\mathbf{r})}$ [Eq.~(\ref{chipm})] of (\ref{H2l}), 
\begin{equation}
|\chi_{ij}(\mathbf{r}_a,\mathbf{r}_b)\rangle=|\chi_{i}(\mathbf{r}_a)\rangle_a\otimes|\chi_{j}(\mathbf{r}_b)\rangle_b,
\label{2atomsindvecp}
\end{equation}
with $i,j=\pm$.

The full Hamiltonian, including quantization of the atomic motion, is given by
\begin{equation}
\hat{H}=\left(\frac{\hat{\mathbf{p}}_a^2}{2m_a}+\frac{\hat{\mathbf{p}}_b^2}{2m_b}+U\right)\otimes\hat{\mathbb{1}}_{ab}^{\mathrm{int}} +\hat{\mathbb{1}}_{ab}^{\mathrm{ext}}\otimes\hat{H}_{\mathrm{non-int}},
\label{2atomsindHtot}
\end{equation}
where $\hat{\mathbb{1}}_{ab}^{\boldsymbol{\cdot}}=\hat{\mathbb{1}}_{a}^{\boldsymbol{\cdot}}\otimes \hat{\mathbb{1}}_{b}^{\boldsymbol{\cdot}}$, $\hat{\mathbf{p}}_\alpha=\hbar\boldsymbol{\nabla}_{\mathbf{r}_\alpha}/i$ is the momentum operator for atom $\alpha$, and $U$ is a potential energy insensitive to the atomic internal state. The global state of the two-atom system is given in the position representation by
\begin{equation}
\langle\mathbf{r}_a,\mathbf{r}_b|\psi(t)\rangle= \sum_{i,j=\pm}{\psi_{ij}(\mathbf{r}_a,\mathbf{r}_b,t) |\chi_{ij}(\mathbf{r}_a,\mathbf{r}_b)\rangle},
\label{2atomsindglobal}
\end{equation}
where the wave functions $\psi_{ij}(\mathbf{r}_a,\mathbf{r}_b,t)$ describe the atomic motion. When the atoms are initially in the (separable) internal eigenstate $|\chi_{ij}(\mathbf{r}_a,\mathbf{r}_b)\rangle$, their adiabatic evolution ensure them to follow the same separable internal state such that $\langle\mathbf{r}_a,\mathbf{r}_b|\psi(t)\rangle\approx\psi_{ij}(\mathbf{r}_a,\mathbf{r}_b,t) |\chi_{ij}(\mathbf{r}_a,\mathbf{r}_b)\rangle$ at any time $t$. Similar developments as before then lead us to the Schr\"{o}dinger-like equation
\begin{equation}\label{2atomsindschroeff}
\begin{aligned}
i\hbar\frac{\partial }{\partial t}\psi_{ij}(\mathbf{r}_a,\mathbf{r}_b,t) =\Bigg[& \sum_{\alpha=a,b}\frac{[\hat{\mathbf{p}}_\alpha-\mathbf{A}_\alpha^{ij}(\mathbf{r}_\alpha)]^2}{2m_\alpha}+\phi_\alpha^{ij}(\mathbf{r}_\alpha)\\
 & +U+E_{i}(\mathbf{r}_a)+E_{j}(\mathbf{r}_b) \Bigg]\psi_{ij}(\mathbf{r}_a,\mathbf{r}_b,t),
\end{aligned}
\end{equation}
with the artificial gauge potentials,
\begin{align}
\mathbf{A}_\alpha^{ij}(\mathbf{r}_\alpha) &=i\hbar\langle \chi_{ij}|\boldsymbol{\nabla}_{\mathbf{r}_\alpha}\chi_{ij}\rangle,\label{2atomsartificielA}\\
\phi_\alpha^{ij}(\mathbf{r}_\alpha) &=\frac{\hbar^2}{2m_\alpha}\sum_{kl\neq ij}{|\langle\chi_{kl}|\boldsymbol{\nabla}_{\mathbf{r}_\alpha}\chi_{ij}\rangle|^2},\label{2atomsartificielphi}
\end{align}
where the sum runs over all eigenstates of the two-atom Hamiltonian except the initial state. For independent atoms, the eigenstates $\ket{\chi_{ij}}$ are separable [Eq.~(\ref{2atomsindvecp})] and Eqs.~(\ref{2atomsartificielA}) and (\ref{2atomsartificielphi}) reduce to the single-atom potentials (\ref{2nivA}) and (\ref{phieff}). As could be expected when the electromagnetic field is treated classically, e.g., as an external field, the noninteracting atoms experience the same artificial gauge potentials as those calculated for a single atom, with the slight difference that they can experience different potentials depending on their respective internal state. Again, when $\delta$ and $\Omega$ are constant over space, the artificial vector potentials are constant and the artificial magnetic fields vanish everywhere.

The generalization of these results to a system of $N$ noninteracting atoms in a classical laser field proceeds along the same lines.

\section{Artificial gauge potentials for two interacting Rydberg atoms}

In order to highlight the contribution of atom-atom interactions on the generation of artificial gauge fields, we consider uniform Rabi frequency and detuning. In this case, the single-atom artificial gauge potentials are constant and do not give rise to any magnetic or electric fields, as recalled in the previous section.

\subsection{Hamiltonian}

We consider a system of two atoms interacting with each other when they are both in an excited Rydberg state. The interaction energy between a ground-state atom and the other atom is assumed to be negligible~\cite{Beg13}. To account for the energy shift of the doubly excited state $|ee\rangle\equiv |e\rangle_a\otimes |e\rangle_b$ caused by dipole-dipole interactions, the term $\hbar V|ee\rangle\langle ee|$ is added to the Hamiltonian (\ref{Hind}) describing two independent atoms driven by a common laser field. This leads us to the Hamiltonian, 
\begin{equation}\label{Hd-d}
\hat{H}_{\mathrm{d-d}}= \hat{H}_{\mathrm{non-int}}+\hbar V |ee\rangle\langle ee|.
\end{equation} 
This simple but realistic model provides the core foundation for several theoretical works on Rydberg gases~\cite{Ate07a}. Gillet \emph{et.~al.}~\cite{Gil10} showed that it successfully reproduces experimental observations on the dipole blockade effect~\cite{Gae09}. Very recently, it was applied by B\'{e}guin and coworkers~\cite{Beg13} to deduce from experimentally measured excitations probabilities the $1/r_{ab}^6$ dependence of the van der Waals interaction between two Rydberg atoms (here $r_{ab}=|\mathbf{r}_a-\mathbf{r}_b|$ is the interatomic distance).

For resonant dipole-dipole (RDD) interactions, the energy shift takes the form $\hbar V =\hbar C_3/r_{ab}^3$. In the absence of an external field, $C_3$ has no angular dependence because of rotational invariance. However, when the atoms are excited by a laser field, $C_3$ may depend on the angle between the interatomic axis and light polarization. Here, we shall consider the case of $s$-state atoms for which the energy shift is almost spherically symmetric~\cite{Saf10}, even though our analysis can be generalized directly to account for an angular dependence. For van der Waals (vdW) interactions, when the atoms are far apart or in the absence of F\"{o}rster resonance, the energy shift displays the characteristic $1/r_{ab}^6$ dependence, $\hbar V =\hbar C_6/r_{ab}^6$. Let us define a crossover distance $r_c$ at which the atom-atom interaction energy equals the atom-field interaction energy. The distance $r_c$ is implicitely determined by the equality
\begin{equation}\label{rc}
|V(r_c)|=\Lambda,
\end{equation}
where $\Lambda$ is given by Eq.~(\ref{Lambda}). For RDD interactions, we have $r_c=\sqrt[3]{|C_3|/\Lambda}$, whereas for vdW interactions, $r_c=\sqrt[6]{|C_6|/\Lambda}$.

\subsection{Eigenstates}
It is convenient to introduce the symmetric and antisymmetric one-excitation states
\begin{equation}
|\psi_{\pm}\rangle= \frac{1}{\sqrt{2}} (e^{i\mathbf{k}_L\boldsymbol{\cdot}\mathbf{r}_{a}}|eg\rangle \pm e^{i\mathbf{k}_L\boldsymbol{\cdot}\mathbf{r}_{b}}|ge\rangle),
\label{2atomspsi+-}
\end{equation}
because $|\psi_-\rangle$ is a trivial eigenstate of $\hat{H}_{\mathrm{d-d}}$ with eigenvalue $0$. In the basis $\{|\psi_-\rangle,|ee\rangle,|\psi_+\rangle,|gg\rangle \}$, Hamiltonian $\hat{H}_{\mathrm{d-d}}$ reads
\begin{equation}\label{Hdd}
\begin{aligned}
\hat{H}_{\mathrm{d-d}}={}& \hbar(V-\delta)|ee\rangle\langle ee|+\hbar\delta|gg\rangle\langle gg|\\
& +\left[\frac{\hbar\Omega}{\sqrt{2}}\left(e^{i\mathbf{k}_L\boldsymbol{\cdot}(\mathbf{r}_{a}+\mathbf{r}_{b})}|ee\rangle\langle \psi_+|+|\psi_+\rangle\langle gg|\right)+\mathrm{h.c.}\right]
\end{aligned}
\end{equation}
It has non-degenerate eigenvalues
\begin{equation}\label{ev}
\begin{aligned}
& E_0=0,\;\;\;\;E_1=\frac{\hbar}{2}\left[s_++s_-+\frac{2}{3}V\right]\\
& E_{\pm}=\frac{\hbar}{2}\left[-\frac{1}{2}\left(s_++s_-\right)+\frac{2}{3}V\pm
i\frac{\sqrt{3}}{2}(s_+-s_-)\right]
\end{aligned}
\end{equation}
where
\begin{equation}
s_\pm=\sqrt[3]{\gamma\pm\sqrt{\eta^3+\gamma^2}}
\end{equation}
and
\begin{equation}
\begin{aligned}
& \eta=\frac{4}{3}\left(\delta(V-\delta)-|\Omega|^2-\frac{V^2}{3}\right)\\
& \gamma=\frac{V}{3}\left(\frac{8}{9}V^2-4\delta(V-\delta)-2|\Omega|^2\right)
\end{aligned}
\end{equation}

The associated eigenvectors are given by ($i=1,\pm$)
\begin{equation}\label{chii}
\begin{aligned}
|\chi_i\rangle= N_i\Big[ & \hbar \Omega\, e^{i\mathbf{k}_L\boldsymbol{\cdot}(\mathbf{r}_{a}+\mathbf{r}_{b})}\mathcal{E}_i\,|ee\rangle
+\sqrt{2}\mathcal{E}_i\mathcal{F}_i\,|\psi_+\rangle\\
& +\hbar \Omega^* \mathcal{F}_i \,|gg\rangle \Big]
\end{aligned}
\end{equation}
where $N_i\equiv N_i(\mathbf{r}_a,\mathbf{r}_b)$ is a normalization constant and 
\begin{equation}\label{EFi}
\mathcal{E}_i(\mathbf{r}_a,\mathbf{r}_b)=E_i-\hbar \delta,\;\;\;\;\mathcal{F}_i(\mathbf{r}_a,\mathbf{r}_b)=E_i+\hbar (\delta-V)
\end{equation}

\subsection{Artificial gauge potentials and fields}

\subsubsection{General expressions}
\begin{figure}
\begin{center}
\includegraphics[width=0.47\textwidth]{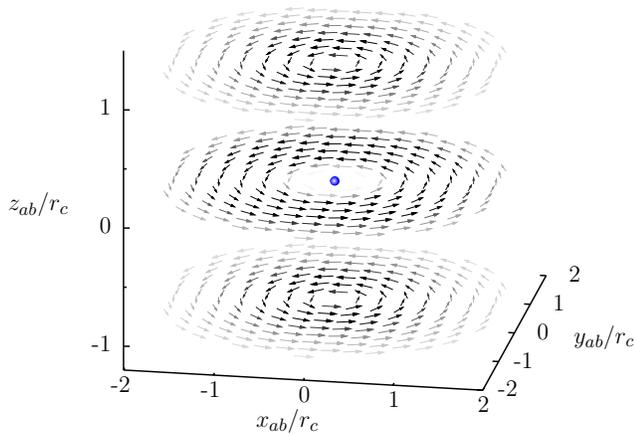}
\end{center}
\caption{(Color online) Vector plot of $\mathbf{B}^+_a$ for $\delta/|\Omega|=0$ and $\mathbf{k}_L=(0,0,k_L)$. Color indicates the magnetic field strength, from black (strong magnetic field) to white (zero magnetic field). The blue dot shows the position of the fixed atom with which atom $a$ interacts.}
\label{Bm_vfield}
\end{figure}

When the system is initially in the internal state $|\chi_i\rangle$ ($i=1,\pm$) [see Eq.~(\ref{chii})], the general expressions of the artificial gauge potentials are still given by Eqs.~(\ref{2atomsartificielA}) and (\ref{2atomsartificielphi}) but with $\ket{\chi_{ij}}$ replaced by the two-atom eigenstates (\ref{chii}). A direct calculation yields
\begin{equation}
\mathbf{A}^{i}_{\alpha}(r_{ab})=A^{i}_{\alpha}(r_{ab})\,\mathbf{e}_{\mathbf{k}_L},
\label{Abd}
\end{equation}
with $r_{ab}=|\mathbf{r}_a-\mathbf{r}_b|$, $\mathbf{e}_{\mathbf{k}_L}=\mathbf{k}_L/k_L$ and
\begin{equation}\label{Aia}
A^{i}_{\alpha}(r_{ab})=-N_i^2\mathcal{E}_i^2\left(\hbar^2|\Omega|^2+\mathcal{F}_i^2\right)\hbar k_L.
\end{equation}
The dependence of $A^{i}_{\alpha}(r_{ab})$ on the position $\mathbf{r}_\alpha$ of atom $\alpha$ appears only through the interatomic distance $r_{ab}$ via $N_i$, $E_i$, and $V$. Since the artificial potentials (\ref{Abd}) are identical for both atoms and depend only on $r_{ab}$, the magnetic fields for atoms $a$ and $b$ have opposite signs. To reduce the amount of notation, we only give the magnetic field experienced by atom $a$, which reads
\begin{equation}
\begin{aligned}
\mathbf{B}^{i}_{a}(\mathbf{r}_{ab}) &{}=\boldsymbol{\nabla}_{\mathbf{r}_a}A^{i}_{a}\times\mathbf{e}_{\mathbf{k}_L}\\
&{}= \frac{dA^{i}_{a}}{dr_{ab}}\,\mathbf{e}_{\mathbf{r}_{ab}}\times\mathbf{e}_{\mathbf{k}_L},
\label{Bbd}
\end{aligned}
\end{equation}
with $\mathbf{e}_{\mathbf{r}_{ab}}=\mathbf{r}_{ab}/r_{ab}$. In a reference frame in which atom $b$ is at the origin, and equipped with spherical coordinates $\{r_{ab},\theta,\varphi\}$ where the $z$-axis points in the same direction as the laser wave vector, the artificial magnetic field takes the form,
\begin{equation}
\begin{aligned}
\mathbf{B}^{i}_{a}(\mathbf{r}_{ab}) =\frac{dA^{i}_{a}}{dr_{ab}}\sin\theta\,\mathbf{e}_\varphi=B^{i}_{a,\varphi}\,\mathbf{e}_\varphi
\label{Bbdphi}
\end{aligned}
\end{equation}
where $\theta$ is the angle between the $z$ axis and $\mathbf{r}_{ab}$. The structure of this vector field is illustrated in Fig.~\ref{Bm_vfield}.

From Eqs.~(\ref{2atomsartificielphi}) and (\ref{chii}), we obtain for the artificial scalar potentials
\begin{equation}\label{aspg}
\begin{aligned}
\phi_\alpha^i(r_{ab})=&N_i^2\bigg\{\frac{\mathcal{E}_i^2\mathcal{F}_i^2}{2}+
\sum_{j\neq i} N_j^2\left[(\mathcal{E}'_i\mathcal{E}_j C_{ij}^{\mathcal{F}}+C_{ij}^{\mathcal{E}}\mathcal{F}'_i\mathcal{F}_j)^2/k_L^2\right.\\
&\left.+\mathcal{E}_i^2\mathcal{E}_j^2(\hbar^2|\Omega|^2+\mathcal{F}_i\mathcal{F}_j)^2
\right]\bigg\}\frac{\hbar^2k_L^2}{2m_\alpha},
\end{aligned}
\end{equation}
with $C_{ij}^{\mathcal{E}}=\hbar^2|\Omega|^2+2\mathcal{E}_i\mathcal{E}_j$, $C_{ij}^{\mathcal{F}}=\hbar^2|\Omega|^2+2\mathcal{F}_i\mathcal{F}_j$, $i,j=1,\pm$ and where the prime denotes a derivative with respect to $r_{ab}$.

Artificial scalar potentials are of the order of the recoil energy $\hbar^2k_L^2/2m$, which for Rubidium atoms and an optical transition is about 1 $\mu$K. This is usually much weaker than the trapping potential $U$ and can be compensated by additional light shifts. We shall therefore concentrate our attention on artificial vector potentials and magnetic fields. For the sake of completeness, we also provide in Appendix A the derivation and a brief discussion of the artificial gauge potentials in the center-of-mass coordinate system.

\subsubsection{Dipole blockade regime}
In the blockade regime, dipole-dipole interactions between excited atoms dominate over atom-light interactions. This prevents the system from populating the doubly excited state $\ket{ee}$, which can be eliminated from the equations of motion, thus leading to an effective Hamiltonian that captures the dynamics as long as the interatomic distance is much smaller than the crossover distance $r_c$ defined by Eq.~(\ref{rc}). The derivation of the effective Hamiltonian and the determination of its eigenvalues and eigenvectors are exposed in Appendix B. There we also show that the artificial gauge potentials take the simple form,
\begin{equation}\label{Aeffdb}
\begin{aligned}
&\mathbf{A}_\alpha^{\mathrm{eff},\pm}(\mathbf{r}_\alpha) =\frac{\mp(\Gamma+\delta)-\sqrt{\Xi}}{2\sqrt{\Xi}}\,\frac{\hbar\mathbf{k}_L}{2},\\[2pt]
&\phi_\alpha^{\mathrm{eff}\pm}(\mathbf{r}_\alpha) = \left[1\pm\frac{\Gamma+\delta}{\sqrt{\Xi}}+\frac{|\Omega|^2}{\Xi}+\frac{4|\Omega|^2|\boldsymbol{\nabla}_{\mathbf{r}_\alpha} \Gamma|^2}{k_L^2\Xi^2}\right]\frac{\hbar^2k_L^2}{16m_\alpha},
\end{aligned}
\end{equation}
where
\begin{equation}
\begin{aligned}
& \Gamma =\frac{|\Omega|^2}{2(V-4\delta/3)},\\[2pt]
& \Xi=(\Gamma+\delta)^2+2|\Omega|^2,
\end{aligned}
\end{equation}
with the correspondence $\mathbf{A}_\alpha^{\mathrm{eff},+}\leftrightarrow \mathbf{A}_\alpha^{+}$, $\mathbf{A}_\alpha^{\mathrm{eff},-}\leftrightarrow \mathbf{A}_\alpha^{-}$ in the case of repulsive interactions and $\mathbf{A}_\alpha^{\mathrm{eff},+}\leftrightarrow \mathbf{A}_\alpha^{-}$, $\mathbf{A}_\alpha^{\mathrm{eff},-}\leftrightarrow \mathbf{A}_\alpha^{1}$ in the case of attractive interactions. The same correspondence holds for the scalar potentials. In this regime, the third vector potential is constant and equal to $-\hbar \mathbf{k}_L$ because the corresponding eigenstate of energy $\hbar (V-\delta)$ reduces to $\exp[i\mathbf{k}_L\boldsymbol{\cdot}(\mathbf{r}_a+\mathbf{r}_b)]\ket{ee}$. For consistency, we checked numerically that these potentials are close to the general expressions (\ref{Abd}) and (\ref{aspg}) for interatomic distances much smaller than $r_c$. 

Expression (\ref{Aeffdb}) for the artificial vector potentials shows a crucial feature: The artificial magnetic fields appear only because of the combined atom-atom and atom-field interactions. Indeed, in the absence of field, $\Omega=0$, the vector potentials are constant and the magnetic fields vanish. On the other hand, when there is no interaction, $V=0$ and the vector potentials are constant which again leads to zero magnetic fields.

For a vanishing detuning ($\delta=0$), the vector potentials become
\begin{equation}
\mathbf{A}_\alpha^{\mathrm{eff},\pm}(\mathbf{r}_\alpha) =\left(-1\mp \frac{|\Omega|}{\sqrt{|\Omega|^2+8V^2(r_{ab})}} \right)\frac{\hbar \mathbf{k}_L}{4}.
\label{Apmdb}
\end{equation}
Comparison of Eq.~(\ref{Apmdb}) with Eq.~(\ref{2leva-phi-}) shows that the artifical vector potentials have the same form, up to a multiplicative factor $1/2$, as those felt by a single two-level atom irradiated by a laser field where $|\Omega|$ plays the role of the detuning and $8V^2(r_{ab})$ the role of the Rabi frequency. This similarity breaks down for the scalar potentials and for both potentials in the presence of a detuning in the interacting two-atom system.

\subsubsection{Weak interaction regime}

For large interatomic distances, $r\gg r_c$, atom-light interaction dominates over atom-atom interactions ($\hbar V\ll \hbar \Lambda$). In this limit, a series expansion of the general expression (\ref{Aia}) yields, after some algebra,
\begin{equation}\label{Awr}
\begin{aligned}
& \mathbf{A}^1_\alpha=\left[\left(-1+\frac{\delta}{\Lambda}\right)+\left(\frac{2\delta^3+3|\Omega|^2(\delta-\Lambda)}{6\Lambda^4}\right)V\right]\frac{\hbar \mathbf{k}_L}{2},\\
& \mathbf{A}^+_\alpha=\left[\left(-1-\frac{\delta}{\Lambda}\right)+\left(\frac{2\delta^3+3|\Omega|^2(\delta+\Lambda)}{6\Lambda^4}\right)V\right]\frac{\hbar \mathbf{k}_L}{2},\\
& \mathbf{A}^-_\alpha=\left[-1-\left(\frac{\delta|\Omega|^2}{3\Lambda^4}\right)V\right]\frac{\hbar \mathbf{k}_L}{2}.
\end{aligned}
\end{equation}
In the absence of atom-atom interactions, the vector potentials $\mathbf{A}^1_\alpha$ and  $\mathbf{A}^+_\alpha$ and the scalar potentials reduce to the single-atom potentials~(\ref{2leva-phi-}) as required. We note that for weak interactions, the artificial vector potentials are linear in the interatomic potential.

\subsection{Discussion}

Resonant dipole-dipole interactions give rise to an attractive or repulsive interaction potential between excited atoms of the form $\hbar V =\hbar C_3/r_{ab}^3$, whereas van der Waals interactions are usually attractive with an interaction potential of the form $\hbar V =\hbar C_6/r_{ab}^6$~\cite{Saf10,comment_branches}. When the laser frequency matches the atomic transition frequency ($\delta=\omega_L-\omega=0$), the sign of the interaction potential does not affect the artificial magnetic fields~\cite{comment}. More generally, it follows from Eq.~(\ref{ev}) that $E_1(V,\delta)=-E_+(-V,-\delta)$ and $E_-(V,\delta)=-E_-(-V,-\delta)$ where $E_i(V,\delta)$ denotes the energy eigenvalue for an interatomic potential $V$ and a detuning $\delta$. These relations, together with Eqs.~(\ref{EFi}),~(\ref{Aia}) and (\ref{Bbdphi}), imply that $\mathbf{B}^1_\alpha(V,\delta)=\mathbf{B}^+_\alpha(-V,-\delta)$ and $\mathbf{B}^-_\alpha(V,\delta)=\mathbf{B}^-_\alpha(-V,-\delta)$. In the remainder, we choose to focus on attractive potentials ($C_3,C_6<0$) both for RDD and vdW interactions, given that the artificial magnetic fields for repulsive interactions can directly be deduced from those for attractive interactions.

Before we discuss the features of the artificial magnetic fields, it is useful to construct from the characteristic length $r_c$ [see Eq.~(\ref{rc})], the laser wave number $k_L$ and the elementary charge $e$, a characteristic magnetic field strength,
\begin{equation}\label{B0}
B_0=\frac{\hbar k_L}{e r_c}.
\end{equation}
As we shall see, $B_0$ gives the typical strength of the artificial magnetic fields induced by the joint atom-laser and atom-atom interactions for a particle with electric charge $q=e$.

\begin{figure}
\begin{center}
\includegraphics[width=0.42\textwidth]{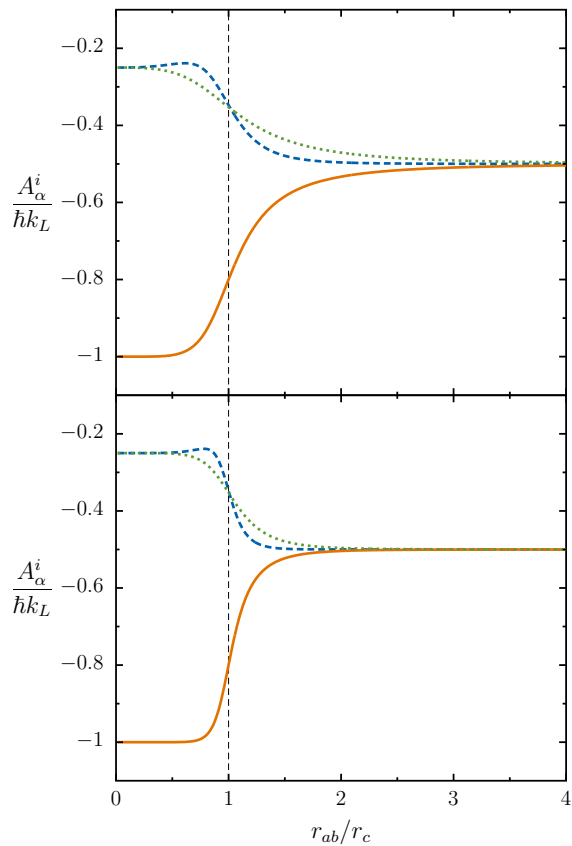}
\end{center}
\caption{(Color online) Only nonvanishing component of the dimensionless artificial vector potentials $\mathbf{A}^i_\alpha/\hbar k_L$ (green dotted curve, $i=1$; blue dashed curve, $i=-$; orange solid curve, $i=+$) as a function of the interatomic distance $r_{ab}/r_c$ for $\delta/|\Omega|=0$, (top panel) resonant dipole-dipole interactions, and (bottom panel) van der Waals interactions.}
\label{Apm1d0fig}
\end{figure}

\begin{figure}
\begin{center}
\includegraphics[width=0.42\textwidth]{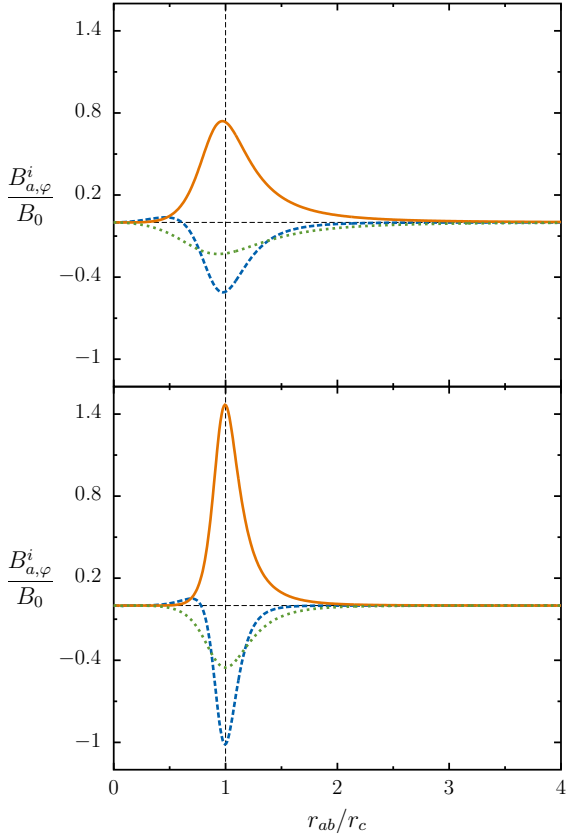}
\end{center}
\caption{(Color online) The $\varphi$-component of the dimensionless magnetic fields $\mathbf{B}^i_a/B_0$ (green dotted curve, $i=1$; blue dashed curve, $i=-$; orange solid curve, $i=+$) as a function of the interatomic distance $r_{ab}/r_c$ for $\delta/|\Omega|=0$, (top panel) resonant dipole-dipole interactions, and (bottom panel) van der Waals interactions.}
\label{Bpm1d0fig}
\end{figure}

We show in Fig.~\ref{Apm1d0fig} the only non-vanishing component of the artificial vector potential as a function of the dimensionless interatomic distance $r_{ab}/r_c$ for a vanishing detuning. Firstly, we note that the artificial vector potentials for RDD and vdW interactions display the same qualitative behavior. For small interatomic distances, $r_{ab}/r_c\ll 1$, the system is dipole blockaded and the component along the laser propagation axis of the vector potential tends to a non-zero value ($-\hbar k_L/4$ or $-\hbar k_L$) which only depends on the atomic internal state, and not on the type of interaction. The largest variations of the vector potentials occur around $r_{ab}=r_c$, and are more pronounced in the case of vdW interactions (see bottom panel). For large $r_{ab}/r_c$, atom-atom interactions become negligible with respect to atom-field interactions, and $A_\alpha^i$ tends in both cases to the value obtained for noninteracting atoms, i.e.\ $-\hbar k_L/2$ [see Eq.~(\ref{2leva-phi-})].
Figure~\ref{Bpm1d0fig} displays the corresponding artificial magnetic fields, which are sizable over a distance interval of the order of $r_c$. A maximum of intensity appears around $r_{ab}= r_c$, where the atom-atom interaction energy $\hbar V$ equals the atom-field interaction energy $\hbar\Lambda$ [see Eq.~(\ref{rc})]. The magnetic field profiles depend markedly on the atomic internal state, and are more squeezed in the case of vdW interactions.

\begin{figure}
\begin{center}
\includegraphics[height=6.5cm]{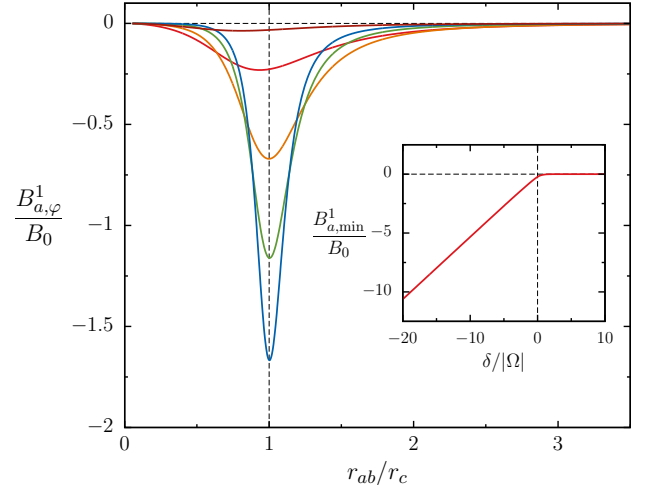}
\end{center}
\caption{(Color online) The $\varphi$-component of the dimensionless magnetic field $\mathbf{B}^1_{a}/B_0$ as a function of the interatomic distance $r_{ab}/r_c$ for different values of the detuning and RDD interactions. (From bottom to top) $\delta/|\Omega|=-3,-2,-1,0,1$. (Inset) Largest value of the artificial magnetic field given by the minimum value of its $\varphi$-component as a function of the dimensionless detuning $\delta/|\Omega|$.}
\label{B1_dsOrL}
\end{figure}

\begin{figure}
\begin{center}
\includegraphics[height=6.5cm]{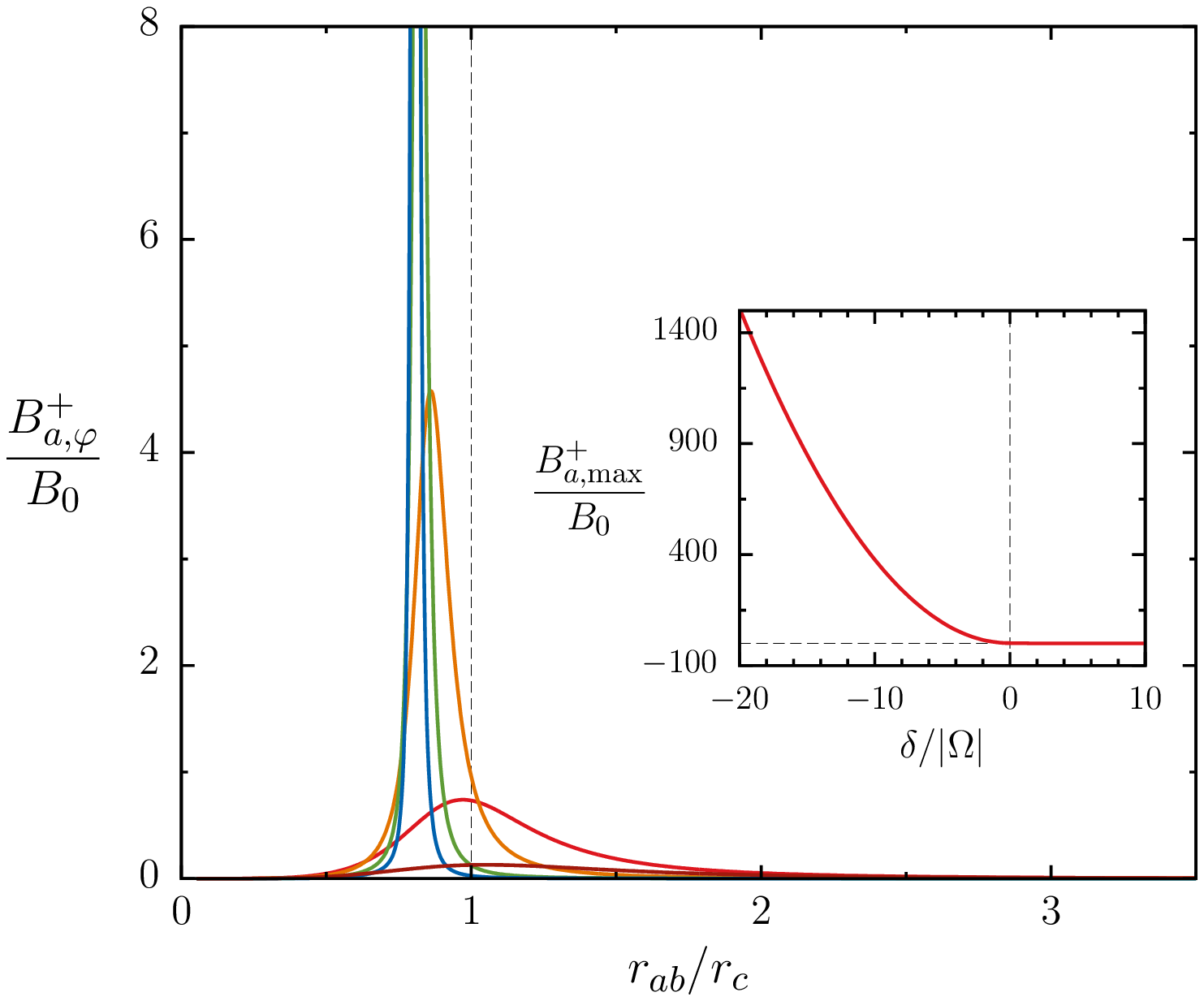}
\end{center}
\caption{(Color online) The $\varphi$-component of the dimensionless magnetic field $\mathbf{B}^+_{a}/B_0$ as a function of the interatomic distance $r_{ab}/r_c$ for different values of the detuning and RDD interactions. (From top to bottom) $\delta/|\Omega|=-3,-2,-1,0,1$. (Inset) Largest value of the artificial magnetic field given by the maximum value of its $\varphi$-component as a function of the dimensionless detuning $\delta/|\Omega|$.}
\label{Bp_dsOrL}
\end{figure}

\begin{figure}
\begin{center}
\includegraphics[height=6.5cm]{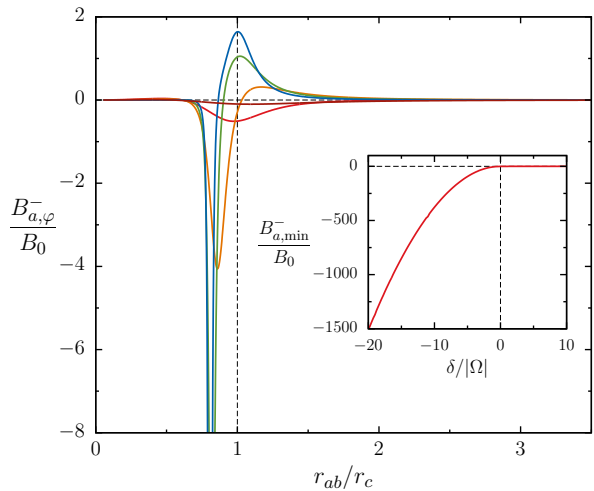}
\end{center}
\caption{(Color online) The $\varphi$-component of the dimensionless magnetic field $\mathbf{B}^-_{a}/B_0$ as a function of the interatomic distance $r_{ab}/r_c$ for different values of the detuning and RDD interactions. (From top to bottom) $\delta/|\Omega|=-3,-2,-1,0,1$. (Inset) Largest value of the artificial magnetic field given by the minimum value of its $\varphi$-component as a function of the dimensionless detuning $\delta/|\Omega|$.}
\label{Bm_dsOrL}
\end{figure}

The behavior of the artificial magnetic field is strongly dependent on the sign of the detuning, as shown in Figs.~\ref{B1_dsOrL}, \ref{Bp_dsOrL} and \ref{Bm_dsOrL} for RDD interactions. The curves for vdW interactions are not shown as they display the same qualitative features. For positive detunings (sign opposite to that of the energy shift $\hbar V$), the overall magnetic field amplitude decreases whereas for negative detunings it increases as compared to the zero detuning case. A series expansion of Eq.~(\ref{Aia}) inserted into Eq.~(\ref{Bbdphi}) shows that for large negative detunings ($\delta<0$ with $|\delta/\Omega|\gg 1$), the peak height of the dimensionless magnetic field $\mathbf{B}^1_a/B_0$ scales linearly with the dimensionless detuning $\delta/|\Omega|$ according to $B^1_{a,\textrm{min}}/B_0\approx \beta^1\,\delta/|\Omega|$ with $\beta^1=3/(4\sqrt{2})$ for RDD interactions and $\beta^1=3/(2\sqrt{2})$ for vdW interactions. Given Eqs.~(\ref{rc}) and (\ref{B0}), it follows that $|B^1_{\textrm{min}}|$ scales like $|\delta|^{4/3}/|\Omega|$ for RDD interactions and like $|\delta|^{7/6}/|\Omega|$ for vdW interactions. This behavior is illustrated in the inset of Fig.~\ref{B1_dsOrL}. The position of the magnetic field peak appears around $r_{ab}= r_c$ as in the zero detuning case, which corresponds to the distance where atom-atom and atom-field interaction energies are equal.

The magnetic field $\mathbf{B}^{+}_a$ displays a single peak around $r_{ab}= \gamma^+\,r_c$, as Fig.~\ref{Bp_dsOrL} shows. For large negative detunings, its intensity scales quadratically with $\delta/|\Omega|$ according to $B^{+}_{a,\textrm{max}}/B_0\approx \beta^+\,(\delta/|\Omega|)^2$ with $\gamma^+=2^{-1/3}\approx 0.79$, $\beta^+=3\sqrt[3]{2}$ for RDD interactions (see inset of Fig.~\ref{Bp_dsOrL}), and $\gamma^+=2^{-1/6}\approx 0.89$, $\beta^+=6\sqrt[6]{2}$ for vdW interactions. 
As regards $\mathbf{B}^-_a$, it displays both a maximum and a minimum (see Fig.~\ref{Bm_dsOrL}). For large negative detunings, the minimum occurs around $r_{ab}= \gamma^-_{\textrm{min}}\,r_c$ and scales quadratically with the detuning, i.e., $B^-_{a,\textrm{min}}/B_0\approx -\beta^-_{\textrm{min}}\,(\delta/|\Omega|)^2$  with $\gamma^-_{\textrm{min}}=2^{-1/3}\approx 0.79$, $\beta^-_{\textrm{min}}=3\sqrt[3]{2}$ for RDD interactions (see inset of Fig.~\ref{Bm_dsOrL}), and $\gamma^-_{\textrm{min}}=2^{-1/6}\approx 0.89$, $\beta^-_{\textrm{min}}=6\sqrt[6]{2}$ for vdW interactions. In the same limit, the maximum occurs around $r_{ab}\approx r_c$, and scales linearly with the detuning according to $ B^-_{a,\textrm{max}}/B_0\approx -\beta^-_{\textrm{max}}\,\delta/|\Omega|$ for $\delta/|\Omega|\gg 1$ with $\beta^-_{\textrm{max}}=3/(4\sqrt{2})$ for RDD interactions and $\beta^-_{\textrm{max}}=3/(2\sqrt{2})$ for vdW interactions. 

The location of the intensity peaks displayed by the magnetic field in the regime of large detunings can be related to transitions between bare states. For negative detunings large compared to the Rabi frequency, transitions between bare states are highly inhibited. However, at small interatomic distances, dipole-dipole interactions give rise to an energy shift of the doubly excited state which can compensate the energy mismatch stemming from the detuning. When the two-photon antiblockade condition $2\hbar\omega+\hbar V=2\hbar\omega_L$ is met, the $\ket{gg}\leftrightarrow\ket{ee}$ transition is on resonance. Similarly, the $\ket{\psi_+}\leftrightarrow\ket{ee}$ transition becomes resonant when the single-photon antiblockade condition $\hbar\omega+\hbar V=\hbar\omega_L$ holds~\cite{Ate07b,Amt10}. In terms of interatomic distances, the two-photon condition reads $r_{ab}=\sqrt[3]{|C_3|/(2\delta)}=r_c\sqrt[3]{\Lambda/(2\delta)}$ for RDD interactions and $r_{ab}=\sqrt[6]{|C_6|/(2\delta)}=r_c\sqrt[6]{\Lambda/(2\delta)}$ for vdW interactions. For large values of the detuning, $\Lambda\approx |\delta|$ such that the condition becomes $r_{ab}\approx r_c/\sqrt[3]{2}\approx 0.79\,r_c$ (RDD) and $r_{ab}\approx r_c/\sqrt[6]{2}\approx 0.89\,r_c$ (vdW). Likewise, the single-photon antiblockade condition reads $r_{ab}\approx r_c$ for both types of interactions. The distances corresponding to these antiblockade conditions coincide with the locations where the magnetic fields are found to be the most intense. This can be understood as follows. Equations~(\ref{2atomsartificielA}) and (\ref{Bbd}) show that large artificial magnetic fields appear where the eigenstates present strong nonuniform spatial variations. This is not the case in the presence of large detunings because transitions between the bare states $\ket{gg}$, $\ket{\psi_+}$, and $\ket{ee}$ are then highly inhibited. However, some of these transitions are enabled at interatomic distances where the antiblockade conditions are met. This is only possible when the detuning and the energy shift $\hbar V$ have the same sign. In this case (negative detunings), the eigenstates display important spatial variations which lead to large artificial magnetic fields, whereas in the absence of antiblockade (positive detunings) the spatial variation of the eigenstates, and thus the magnetic fields, are small (see~insets of Figs.~\ref{B1_dsOrL}, \ref{Bp_dsOrL} and \ref{Bm_dsOrL}). Indeed, in the case of $\mathbf{B}^1_a$, the location of the intensity peak ($r_{ab}\approx r_c$) satisfies to the single-photon antiblockade condition. As regards the corresponding eigenstate $\ket{\chi_1}$, it coincides nearly with $\ket{\psi_+}$ at small interatomic distances, turns into a superposition of $\ket{\psi_+}$ and $\ket{ee}$ around $r_{ab}\approx r_c$, and coincides nearly with $\ket{ee}$ at larger distances.
The same observation holds for the artificial magnetic field $\mathbf{B}^+_a$, which displays a maximum of intensity around $r_{ab}\approx r_c/\sqrt[3]{2}$ (RDD) or $r_{ab}\approx r_c/\sqrt[6]{2}$ (vdW) satisfying the two-photon antiblockade condition. In this case, $\ket{\chi_+}$ reduces to $\ket{ee}$ for $r_{ab}\ll r_c$ and to $\ket{gg}$ for $r_{ab}\gg r_c$. When the two-photon antiblockade condition is met, $\ket{\chi_+}$ becomes an equally weighted superposition of $\ket{ee}$ and $\ket{gg}$.
As for the artificial magnetic field $\mathbf{B}^-_a$, the two antiblockade conditions are successively met as the interatomic distance grows because the associated eigenstate coincides with $\ket{gg}$ for $r_{ab}\ll r_c$. The first (second) maximum of intensity corresponds to the two-photon (single-photon) antiblockade condition. At interatomic distances in-between the two maxima, $\ket{\chi_-}$ coincides nearly with $\ket{ee}$, and at large distances with $\ket{\psi_+}$. At the locations of the maxima, $\ket{\chi_-}$ is a superposition of the two states involved in the antiblockade mechanism.

This mechanism also allows one to understand qualitatively the width of the intensity peaks in the artificial magnetic fields. Indeed, the larger the detuning, the smaller the crossover distance $r_c$ and the more important the spatial variations of the dipole shift $\hbar V$ around $r_{ab}=r_c$. Therefore, the interval of distances where the dipole shift and the detuning counterbalance to allow significant transitions between bare states becomes narrower as the detuning increases.

A similar line of reasoning can be pursued to explain the cause of the magnetic field peaks and their location ($r_{ab}\approx r_c$) when the system is at resonance ($\delta=0$). Indeed, at small interatomic distances, the dipole blockade effect prevents the system from populating the doubly excited state, which as a matter of fact do not contribute to the artificial magnetic field. When the dipole shift is of the order of the atom-light interaction energy, transitions between $\ket{ee}$ and lower excitations states are enabled, resulting in spatial variations of the eigenstates leading to artificial magnetic fields. At large distances, the atoms can be considered as independent and the magnetic fields vanish. 

\subsection{Experimental considerations}

In this section, we give an estimation of the attainable artificial magnetic field strengths in the RDD and vdW regimes in view of recent experiments. Moreover, we show that the artificial gauge potentials could be detected through the deflection of the atomic motion caused by the artificial Lorentz force. To verify our theoretical predictions, similar experimental setups as those designed by Ga\"{e}tan~\textit{et.~al.}~\cite{Gae09} (RDD regime) or B\'{e}guin~\textit{et.~al.}~\cite{Beg13} (vdW regime) could be considered. In those experiments, two $^{87}$Rb atoms with residual temperature $T\approx 50$ $\mu$K ($v_{\textrm{rms}}\approx 12$~cm/s) are trapped in two optical tweezers with a beam waist $w\approx 1~\mu$m. The atoms are laser excited to Rydberg states with high principal quantum number ($n=53,62,82$ in~\cite{Beg13} and $n=58$ in~\cite{Gae09}) characterized by a radiative lifetime $\tau$ ranging from $200$ to $500$ $\mu$s. Depending on the principal quantum number, either the RDD or the vdW regime can be reached.

Let us now consider that one atom is kept at a fixed position in space and a second atom is sent towards the first one, e.g., by means of an optical conveyor belt~\cite{Sch01}. The results of the preceding sections show that the moving atom will experience artificial gauge fields as a result of its joint interaction with the trapped atom and the laser field. As a consequence, its trajectory will be modified by the action of the artificial Lorentz force.
In the RDD regime, we consider $|\Omega|/2\pi =6.5$ MHz, $\lambda_L=296$ nm~\cite{comment2}, and $C_3/2\pi=3200$ MHz.$\mu$m$^3$ as in~\cite{Gae09}. In this case, the crossover distance at zero detuning is $r_c\approx 8~\mu$m and the characteristic magnetic field strength $B_0\approx 2$ mT for a particle with electric charge equal to the elementary charge. In the vdW regime, we base our estimation on Ref.~\cite{Beg13} in which the single-atom Rabi frequency $|\Omega|/2\pi$ can be varied in the range from 500 kHz to 5 MHz, and $|C_6|$ can be varied from $10$ to $10000$ GHz.$\mu$m$^6$ by changing the principal quantum number $n$ of the atomic Rydberg state. In this case, the crossover distance at zero detuning can be tuned from $r_c\approx 3.5~\mu$m to $r_c\approx 16~\mu$m and the characteristic magnetic field strength from $B_0\approx 4$ mT to $B_0\approx 0.8$ mT. Note that in both regimes the crossover distance is much larger than the waist of the tweezers.

For an initial velocity of $10$ cm/s in the $xy$ plane and an impact parameter equal to $r_c\approx 8~\mu$m, a semiclassical calculation predicts a deflection of the atomic trajectory in the $z$ direction (laser propagation direction) of the order of 1 $\mu$m for a traveled distance equal to $2 r_c$ in the $xy$ plane. For such a velocity, the adiabatic approximation is still valid to about $99\%$~\cite{Dal11} and it takes a time equal to $160$ $\mu\mathrm{s}<\tau$ to travel a distance $2 r_c$, during which spontaneous emission can be neglected to a good approximation.
This approximation holds even better if we consider that the system follows adiabatically the internal state $\ket{\chi_+}$ corresponding to the two atoms in their ground state at large interatomic distances with respect to $r_c$. In this situation, the Rydberg states are populated only during a small time in comparison with their radiative lifetime $\tau$. Larger (smaller) initial velocities would lead to smaller (larger) deflections. The main experimental challenge is thus to control the atomic velocity with sufficient precision to avoid a drift due to an initial velocity in the $z$ direction that would mask the deflection due to the artificial Lorentz force.

\section{Conclusion}
We have shown that dipole-dipole interactions between Rydberg atoms submitted to a uniform laser field give rise to nonuniform artificial Abelian gauge potentials. We have obtained general analytical expressions for the latter, as well as approximate expressions in the dipole blockade and weak interaction regimes. We have identified the mechanism responsible for the artificial gauge fields and have shown that they are the strongest when atom-atom and atom-field interaction energies are of the same order of magnitude. Note that a similar feature has been observed experimentally in the population dynamics of a pair of interacting Rydberg atoms~\cite{Beg13}. We have discussed the differences and similarities of artificial gauge fields originating from resonant dipole-dipole and van der Waals interactions. We have estimated on the basis of recent experiments the attainable artificial magnetic field to a few mT extending over a distance range of a few micrometers. Finally, we have shown that these fields lead under realistic conditions to a deflection of the atomic motion of the order of 1 $\mu$m, measurable with current imaging techniques~\cite{She10}.

\begin{acknowledgments} 
J.M. is grateful to the University of Li\`{e}ge (SEGI facility) for the use of the NIC3 supercomputer.
\end{acknowledgments}

\section{Appendix A : Artificial gauge fields in the center-of-mass reference frame}

In this appendix, we give the expressions of the artificial electromagnetic potentials induced by dipole-dipole interactions between two Rydberg atoms in the center of mass reference frame. This is the most natural reference frame in view of the interaction potential which only depends on the relative coordinate. We first recall the center-of-mass coordinates,
\begin{equation}
\left\{
\begin{aligned}
\mathbf{R}&=\frac{m_a\mathbf{r}_a+m_b\mathbf{r}_b}{m_a+m_b},\\
\mathbf{r}&=\mathbf{r}_a-\mathbf{r}_b
\end{aligned}
\right.
\label{2atomsrcm}
\end{equation}and their conjugate momenta,
\begin{equation}
\left\{
\begin{aligned}
\mathbf{P}&=\mathbf{p}_a+\mathbf{p}_b,\\
\mathbf{p}&=\frac{m_b\mathbf{p}_a-m_a\mathbf{p}_b}{m_a+m_b}
\end{aligned}
\right.
\label{2atomspcm}
\end{equation}where $m_\alpha$, $\mathbf{r}_\alpha$ and, $\mathbf{p}_\alpha$ are, respectively, the mass, the position and the momentum of atom $\alpha=a,b$. The total mass of the system is $M=m_a+m_b$ and the reduced mass is $\mu=m_a m_b/(m_a+m_b)$. In the center-of-mass reference frame, the full Hamiltonian takes the form,
\begin{equation}
\hat{H}=\left(\frac{\hat{\mathbf{P}}^2}{2M}+\frac{\hat{\mathbf{p}}^2}{2\mu}+U\right)\otimes\hat{\mathbb{1}}_{ab}^{\mathrm{int}} +\hat{\mathbb{1}}_{ab}^{\mathrm{ext}}\otimes\hat{H}_{\mathrm{d-d}}(\mathbf{R},\mathbf{r}),
\label{2atomsindHtot}
\end{equation}
where $\hat{\mathbf{P}}=\hbar\boldsymbol{\nabla}_{\mathbf{R}}/i$ and $\hat{\mathbf{p}}=\hbar\boldsymbol{\nabla}_{\mathbf{r}}/i$ in the position representation, and with $\hat{H}_{\mathrm{d-d}}(\mathbf{R},\mathbf{r})$ given by Eq.~(\ref{Hd-d}) but now expressed in terms of the center-of-mass coordinates (\ref{2atomsrcm}). The eigenstates of $\hat{H}_{\mathrm{d-d}}(\mathbf{R},\mathbf{r})$ are still given by Eq.~(\ref{chii}) and, following the same procedure as in Sec. II, we obtain 
\begin{equation}
\begin{aligned}
\mathbf{A}_\mathbf{R}^{i}(\mathbf{r}) &=\mathbf{A}_a^{i}+\mathbf{A}_b^{i},\\
\mathbf{A}_\mathbf{r}^{i}(\mathbf{r}) &=\frac{m_b\mathbf{A}_a^{i}-m_a\mathbf{A}_b^{i}}{m_a+m_b}.
\label{2atomsartificielCM}
\end{aligned}
\end{equation}
The resulting artificial vector potentials have the same form as the relations (\ref{2atomspcm}) between the center of mass and relative momenta, and the momenta of atoms $a$ and $b$. This fully agrees with the reformulation of the artificial vector potentials as the expectation values of the momentum operator evaluated in the atomic internal states as presented in Sec. II. In the center-of-mass reference frame, the scalar potentials are given by
\begin{equation}
\begin{aligned}
\phi_{\mathbf{R}}^i(\mathbf{r})=&\sum_{j\neq i}N_i^2N_j^2\left[\mathcal{E}_i^2\mathcal{E}_j^2\right(\hbar^2|\Omega|^2+\mathcal{F}_i\mathcal{F}_j)^2]\frac{\hbar^2k_L^2}{2M},\\
\phi_{\mathbf{r}}^i(\mathbf{r})=&N_i^2\bigg\{\frac{\mathcal{E}_i^2\mathcal{F}_i^2}{2}+
\sum_{j\neq i} N_j^2\left[(\mathcal{E}'_i\mathcal{E}_j C_{ij}^{\mathcal{F}}+C_{ij}^{\mathcal{E}}\mathcal{F}'_i\mathcal{F}_j)^2/k_L^2\right.\\
&\left.+\left(\frac{m_b-m_a}{M}\right)^2\mathcal{E}_i^2\mathcal{E}_j^2(\hbar^2|\Omega|^2+\mathcal{F}_i\mathcal{F}_j)^2
\right]\bigg\}\frac{\hbar^2k_L^2}{2\mu}
\label{2atomsartificielCMphi}
\end{aligned}
\end{equation}
where $C_{ij}^{\mathcal{E}}=\hbar^2|\Omega|^2+2\mathcal{E}_i\mathcal{E}_j$, $C_{ij}^{\mathcal{F}}=\hbar^2|\Omega|^2+2\mathcal{F}_i\mathcal{F}_j$, $i,j=1,\pm$ and where the prime denotes a derivative with respect to $r_{ab}$. They correspond to $(\Delta \hat{\mathbf{P}}^2)_{\chi_i}/2M$ and $(\Delta \hat{\mathbf{p}}^2)_{\chi_i}/2\mu$, respectively, and are thus determined by the variance of center of mass and relative momenta in the two-atom internal state $\ket{\chi_i}$ [see Eq.~(\ref{chii})].

Expressions~(\ref{2atomsartificielCM}) and~(\ref{2atomsartificielCMphi}) show that the potentials in the center-of-mass reference frame are simply connected to those in the laboratory frame. Moreover, when one atom is kept at a fixed position in space and another atom is traveling around it, the vector potential for the relative coordinate reduces to the one for the moving atom in the laboratory frame.

\section{Appendix B : derivation of an effective Hamiltonian in the dipole blockade regime}

In this appendix, we derive an effective Hamiltonian describing the internal dynamics of two interacting Rydberg atoms in the dipole blockade regime. We also determine its eigenvalues and eigenvectors. For this purpose, we eliminate the doubly excited state $|ee\rangle$ from Hamiltonian~(\ref{Hdd}) following a method recently proposed by Paulisch \textit{et al.}~\cite{Pau13}. For the effective Hamiltonian to be valid in the largest possible range, we add, beforehand, a constant term $C \hat{\mathbb{1}}^{\mathrm{int}}_{ab}$ to $\hat{H}_{\mathrm{d-d}}$. This term does obviously not affect the dynamics of the system but leads to a different effective Hamiltonian. Following~\cite{Pau13}, $C$ is chosen so as to satisfy the condition $\mathrm{Tr}(\hat{H}' + C \hat{\mathbb{1}}_3)=0$ where $\hat{H}'$ is the restriction of $\hat{H}_{\mathrm{d-d}}$ to the subspace spanned by $\{|\psi_-\rangle,|\psi_+\rangle,|gg\rangle \}$ and $\hat{\mathbb{1}}_3$ the identity operator in this subspace. A straightforward calculation shows that $C=-\hbar \delta/3$. We now eliminate the state $|ee\rangle$ from Hamiltonian $\hat{H}_{\mathrm{d-d}}'=\hat{H}_{\mathrm{d-d}}-(\hbar \delta/3)\hat{\mathbb{1}}^{\mathrm{int}}_{ab}$ by first writing the internal states of the two-atom system in the form
\begin{equation}
|\psi(t)\rangle =c_e(t)|ee\rangle+c_+(t)|\psi_+\rangle+c_-(t)|\psi_-\rangle+c_g(t)|gg\rangle.
\end{equation}
Inserting this expression into the time-dependent Schr\"{o}dinger's equation, we obtain the set of equations:
\begin{eqnarray}
i\,\dot{c}_e&=&\Delta \,c_e+\frac{\Omega}{\sqrt{2}}e^{i\mathbf{k}_L\boldsymbol{\cdot}(\mathbf{r}_a+\mathbf{r}_b)}\,c_+\label{eq_ce},\\
i\,\dot{c}_+&=&-\frac{\delta}{3}\,c_++\frac{\Omega}{\sqrt{2}}c_g+\frac{\Omega^*}{\sqrt{2}}e^{-i\mathbf{k}_L\boldsymbol{\cdot}(\mathbf{r}_a+\mathbf{r}_b)}\,c_e\label{eq_c+},
\\
i\, \dot{c}_-&=&-\frac{\delta}{3}\,c_-,\\
i\, \dot{c}_g&=&\frac{2\delta}{3}\,c_g+\frac{\Omega^*}{\sqrt{2}}\,c_+\label{eq_cg},
\end{eqnarray}
where a dot denotes a time derivative and $\Delta=V-4\delta/3$. Solving Eq.~(\ref{eq_ce}) for $c_e(t)$, we get
\begin{equation}
c_e(t)=-\frac{i}{\sqrt{2}}\int_0^t{e^{-i\Delta(t-t')}\Omega e^{i\mathbf{k}_L\boldsymbol{\cdot}(\mathbf{r}_a+\mathbf{r}_b)}c_+(t')dt'}.
\end{equation}
In the Markov approximation, memory effects are neglected, which amounts to taking the coefficient $c_+(t')$ out of the integral. This approximation is valid as long as $\Lambda=\sqrt{\delta^2+|\Omega|^2}\ll|V|$ and implies that $c_i(t)$ ($i\neq e$) oscillates slowly in comparison to $\exp{(-iVt)}$. In this case, $c_e(t)$ takes the simple form,
\begin{equation}
c_e(t)=-\frac{\Omega}{\sqrt{2} \Delta}e^{i\mathbf{k}_L\boldsymbol{\cdot}(\mathbf{r}_a+\mathbf{r}_b)}c_+(t).
\end{equation}
Inserting this expression into Eq.~(\ref{eq_c+}), we readily deduce from the equations of motion (\ref{eq_c+})--(\ref{eq_cg}) the effective Hamiltonian,
\begin{equation}\label{2atomsintHddeff}
\begin{aligned}
\hat{H}_{\mathrm{d-d}}^{\mathrm{eff}}={}&-\frac{\hbar}{3}(3\Gamma+\delta)\ket{\psi_+}\bra{\psi_+}\\
& +\frac{\hbar\delta}{3}(2\ket{gg}\bra{gg}-\ket{\psi_-}\bra{\psi_-})\\
& + \left(\frac{\hbar\Omega}{\sqrt{2}}\ket{\psi_+}\bra{gg}+\mathrm{h.c.}\right)
\end{aligned}
\end{equation}
where $\Gamma =|\Omega|^2/2\Delta$. The state $|\psi_-\rangle\equiv \ket{\chi_0}$ remains eigenstate of the effective Hamiltonian, but with energy $E_0=-\hbar \delta/3$. The two other eigenstates are
\begin{equation}
\begin{aligned}
|\chi_\pm^{\mathrm{eff}}\rangle= N_{\pm}\bigg[{}&\left(-(\Gamma+\delta)\mp\sqrt{(\Gamma+\delta)^2+2|\Omega|^2}\,\right)|\psi_+\rangle\\
& +\sqrt{2}\Omega^*\,|gg\rangle\bigg]
\label{2atomsintvecp!0}
\end{aligned}
\end{equation}
with eigenvalues
\begin{equation}
E_\pm^{\mathrm{eff}}=\frac{\hbar}{6}\left[\delta-3\Gamma\mp 3\sqrt{(\Gamma+\delta)^2+2|\Omega|^2}\,\right].
\label{2atomsintEp!0}
\end{equation}

Similarly to Eqs.~(\ref{2atomsartificielA}) and (\ref{2atomsartificielphi}) of Sec. II, the artificial gauge potentials are given by
\begin{equation}
\begin{aligned}
\mathbf{A}_\alpha^{\mathrm{eff},\pm}(\mathbf{r}_\alpha) &= i\hbar\langle \chi_{\pm}^{\mathrm{eff}}|\boldsymbol{\nabla}_{\mathbf{r}_\alpha}\chi_{\pm}^{\mathrm{eff}}\rangle,\\
\phi_\alpha^{\mathrm{eff},\pm}(\mathbf{r}_\alpha) &=\frac{\hbar^2}{2m_\alpha}\left({|\langle\chi_{\mp}^{\mathrm{eff}}|\boldsymbol{\nabla}_{\mathbf{r}_\alpha}\chi_{\pm}^{\mathrm{eff}}\rangle|^2+|\langle\chi_{0}|\boldsymbol{\nabla}_{\mathbf{r}_\alpha}\chi_{\pm}^{\mathrm{eff}}\rangle|^2}\right).
\end{aligned}
\end{equation}
After some algebra, we arrive at Eq.~(\ref{Aeffdb}).
When the signs of the detuning ($\delta$) and the interatomic potential ($V$) are changed simultaneously, $E_\pm^{\mathrm{eff}}\to -E_\mp^{\mathrm{eff}}$, and $\mathbf{A}_\alpha^{\mathrm{eff},\pm}\to \mathbf{A}_\alpha^{\mathrm{eff},\mp}$.


\begin{thebibliography}{99}
\bibitem{comment_branches} Note that in our model, the sign of the interatomic potential is fixed.
\bibitem{comment} However, a change of sign of the interaction potential at zero detuning turns $\mathbf{B}^1_\alpha$ into $\mathbf{B}^+_\alpha$ and vice versa, whereas $\mathbf{B}^-_\alpha$ remains unchanged.
\bibitem{comment2} It should be mentioned that the experiment exploits a two-photon transition driven by lasers with wavevectors $\mathbf{k}_1$ and $\mathbf{k}_2$. The resulting wavevector $\mathbf{k}_L=\mathbf{k}_1+\mathbf{k}_2$ leads to an effective wavelength $\lambda_L=2\pi/k_L$.
\end{thebibliography}
\end{document}